\RequirePackage{fix-cm}
\documentclass[11pt]{article}
\usepackage[margin=1in]{geometry}
\usepackage{graphicx}
\usepackage{graphics} 
\usepackage{mathptmx} 
\usepackage{times} 
\usepackage{amsmath} 
\usepackage{amssymb}  
\usepackage{amsfonts,bm} 
\usepackage{amsthm} 
\usepackage{algorithm,algpseudocode}
\usepackage{mathrsfs}
\usepackage{blkarray, bigstrut}
\DeclareMathAlphabet{\mathcal}{OMS}{cmsy}{m}{n}
\usepackage{graphicx}
\usepackage{pstool}
\usepackage{hyperref}
\usepackage{dblfloatfix}    
\usepackage{enumitem}

\usepackage{multirow}
\usepackage{makecell}
\usepackage{tabularx} 

\usepackage{pgfplots}
\pgfplotsset{compat=newest}
\usetikzlibrary{plotmarks}
\usetikzlibrary{arrows.meta}
\usepgfplotslibrary{patchplots}
\usepackage{grffile}
\usepackage{amsmath}
\usepackage{color,soul} 

\DeclareMathOperator{\real}{\mathbb{R}}
\DeclareMathOperator{\likelihood}{\mathcal{L}}

\DeclareMathOperator{\tangentspace}{\mathcal{T}_x\mathcal{S}}
\newtheorem{problem}{Problem}
\newtheorem{definition}{Definition}
\newtheorem{proposition}{Proposition}
\theoremstyle{definition}
\newtheorem{example}{Example}
\newtheorem{experiment}{Experiment}

\newcommand{\norm}[1]{\left\lVert#1\right\rVert}

\usepackage{marvosym} 

\makeatletter
\providecommand{\keywords}[1]{%
	\par\smallskip\noindent\textbf{Keywords. }#1\par\smallskip}
\providecommand{\subclass}[1]{%
	\par\smallskip\noindent\textbf{MSC (2020). }#1\par\smallskip}
\makeatother

\title{\LARGE \bf
	Probability-One Optimization of Generalized Rayleigh Quotient Sum For Multi-Source Generalized Total Least-Squares
}

\author{Dominik Friml 
	\thanks{Dominik Friml is with the~Central European Institute of Technology, Purkyňova 656/123b Brno, Czech Republic
		{\tt\small dominik.friml@vutbr.cz, +420 54114 9823}}%
	\and Pavel Václavek
	\thanks{Pavel Václavek is with the~Faculty of Electrical Engineering and Communication, 
		Brno University of Technology, and with the~Central European Institute of Technology, Technická 12 Brno, Czech Republic
		{\tt\small pavel.vaclavek@vutbr.cz, +420 54114 6413}}%
}

\begin{document}
	
	\title{Probability-One Optimization of Generalized Rayleigh Quotient Sum For Multi-Source Generalized Total Least-Squares
	}
	
	
	
	
	
	\date{}
	
	\maketitle

	
	\begin{abstract}
		This paper addresses the global optimization of the sum of the Rayleigh quotient and the generalized Rayleigh quotient on the unit sphere. While various methods have been proposed for this problem, they fail to reliably converge to the global maximizer. To overcome this limitation, we propose an extension of the Riemannian Trust Region algorithm based on the probability-one homotopy optimization method, which enhances convergence to a global maximizer and, under certain conditions, ensures convergence to the global maximizer. In addition to the proposed method, existing state-of-the-art approaches are also presented, along with an explanation of their limitations and their connection to the proposed method. The proposed method is evaluated alongside the state-of-the-art approaches through numerical experiments, assessing convergence speed, success in reaching the global maximizer, and scalability with increasing problem dimension. Furthermore, we demonstrate how this ties in with the multi-source Bayesian Generalized Total Least-Squares (B-GTLS) problem, illustrating its applicability.
		
		\subclass{90C26 \and 65H20 \and 65K10 \and 90C48}
	\end{abstract}

	\section{Introduction}
	
	We consider the problem of finding a global maximizer for a nonlinear, antipodally symmetric objective function composed of a Rayleigh quotient and a generalized Rayleigh quotient, constrained to the unit sphere. 
	\begin{problem}\label{prob:problem}
		Given symmetric matrices \(B, D \in \mathbb{R}^{q \times q}\) and a symmetric positive definite matrix \(W \in \mathbb{R}^{q \times q}\), find a vector \(x^* \in \mathbb{R}^q\) that solves the optimization problem
		\begin{align}\label{eq:dualOptimizationProblem}
			\begin{split}
				x^* := \operatorname*{arg\,max}_{x \in \mathbb{R}^q}  &\quad \frac{x^T B x}{x^T W x} + x^T D x. \\
				\textrm{s.t.} &\quad \|x\| = 1
			\end{split}
		\end{align}
	\end{problem}
	
	Depending on the properties of $B$, $D$, and $W$, the maximization problem may be either trivial or nontrivial. Trivial instances are such combinations of parameters $B$, $D$, and $W$ where gradient ascent on the sphere converges to the global maximizer (or maximizer set) from all initializations, except for a set of initializations of measure zero (with respect to the uniform measure). Typically, this measure-zero set consists of stable manifolds of non-maximizer critical points such as local minima or saddle points. Nontrivial instances are all parameter combinations that are not trivial in the above sense. In such problems, optimization can be problematic, because the algorithms may converge to non-global critical points, most notably spurious local maximizers. The goal of this paper is to develop a method that reliably obtains global optimizer $x^*$ in both trivial and nontrivial instances of Problem~\ref{prob:problem}.
	
	This optimization problem arises in several practical applications, including the downlink design of multi-user MIMO systems \cite{primolevo_towards_2006}, sparse Fisher discriminant analysis \cite{zhang_optimizing_2013}, and portfolio optimization \cite{eberhard_generalized_2005}. A novel application in the context of multi-source Bayesian Generalized Total Least-Squares methods is presented in Section~\ref{sec:application}.
	
	Over the years, various strategies for solving Problem~\ref{prob:problem} have been developed. Zhang \cite{zhang_optimizing_2013} proposed a Riemannian trust-region method enhanced by a starting-point strategy. Building on his characterization of the solution, nonlinear eigenvalue approaches emerged \cite{zhang_self-consistent-field-like_2014,zhang_nonlinear_2021}. In \cite{nguyen_maximizing_2016}, a semidefinite programming approach was proposed, which was further refined in \cite{xia_minimizing_2018,wang_linear-time_2019}. The aforementioned approaches are discussed in detail in the following sections. However, all these methods remain imperfect, as the global optimizer in the nontrivial instance is not reliably attained.
	
	In this work, we build upon an existing Riemannian Trust Region optimization framework to develop a homotopy-based probability-one global optimization method, which demonstrably outperforms existing approaches in attaining the global maximizer across both trivial and nontrivial instances. This method is compared with existing methods via numerical experiments.
	
	The paper is structured as follows: Section~\ref{sec:riemannian_analysis} presents the Riemannian analysis and the Riemannian trust-region algorithm. Section~\ref{sec:proposed_method} builds on this by proposing the homotopy method, with its probability-one conditions discussed in Section~\ref{sec:probability_one_conditions}. Section~\ref{sec:stateoftheart_methods} reviews state-of-the-art methods, highlighting their advantages and limitations. Section~\ref{sec:numerical_analysis} conducts numerical analysis, Section~\ref{sec:application} introduces the practical application and the final section concludes the paper.
	
	In this paper, $\norm{\bullet}$ denotes $\ell_2$ norm, local maximizer is denoted $\hat \bullet$, global maximizer is $\bullet^*$; $\bullet^T$ is matrix transpose and $\bullet^{-T}$ is inverse of matrix transpose.

	\section{Riemannian Analysis}\label{sec:riemannian_analysis}
	The constrained optimization problem \eqref{eq:dualOptimizationProblem} can be reformulated as an unconstrained Riemannian optimization problem, that inherently respect the problem's constraint. Considering the unit sphere \( \mathcal{S} := \{x \in \mathbb{R}^q \mid \|x\| = 1\} \) as a submanifold of the Euclidean space \( \mathcal{M} := \mathbb{R}^q \), let \( f:\mathcal{S} \rightarrow\mathbb{R} \) be the smooth Riemannian objective function:
	\begin{align}\label{eq:definition_of_f}
		f(x):=\frac{\phi_b(x)}{\phi_w(x)}+{\phi_d(x)},
	\end{align}
	where \( \phi_b(x) := x^T B x \), \( \phi_w(x) := x^T W x \), and \( \phi_d(x) := x^T D x \). Then, without loss of generality, the optimization problem \eqref{eq:dualOptimizationProblem} becomes 
	\begin{align}\label{eq:RiemannianOptimization}
			x^*:= \operatorname*{arg\,max}_{x\in\mathcal{S} }  &\quad f(x).
	\end{align}
	The Euclidean constrained formulation~\eqref{eq:dualOptimizationProblem} and its Riemannian counterpart~\eqref{eq:RiemannianOptimization} are mathematically equivalent. Accordingly, we will use the notation and viewpoint of either formulation interchangeably, depending on context, as both solve Problem~\ref{prob:problem}.
	
	For \eqref{eq:RiemannianOptimization}, the triviality can be formally defined as:
	\begin{definition}\label{def:triviality}
		Let \( \mathcal{S} := \{x \in \mathbb{R}^q \mid \|x\| = 1\} \), and let \( f : \mathcal{S} \to \mathbb{R} \) be a smooth, antipodally symmetric function. 
		
		We call \eqref{eq:RiemannianOptimization} a trivial maximization problem if:
		\begin{enumerate}
			\item The set of global maximizers $ \mathcal{X}^* \subset \mathcal{S} $ is either a singleton (up to antipodal symmetry) or a connected, antipodally symmetric subset of the sphere.
			\item Consider the Riemannian gradient flow on the manifold \( \mathcal{S} \),
			$$
			\frac{d}{dt} x(t) = \nabla_{\mathcal{S}} f(x(t)), \quad t \in \mathbb{R}_{\geq 0},
			$$
			where \( \nabla_{\mathcal{S}} f(x) \) denotes the Riemannian gradient of \( f \) at \( x \), i.e., the projection of the Euclidean gradient \( \nabla f(x) \) onto the tangent space \( T_x\mathcal{S} \). Then, for almost every initial condition \( x(0) \in \mathcal{S} \), the trajectory \( x(t) \) converges to the maximizer set \( \mathcal{X}^* \), except for a set of initial conditions of measure zero with respect to the uniform measure on \( \mathcal{S} \).
			
		\end{enumerate}
	\end{definition}
	
	It should be noted, that no analytical relationship is known that would directly link $B$, $D$, and $W$ to aforementioned triviality. 
	
	To efficiently solve \eqref{eq:RiemannianOptimization}, we employ the Riemannian Trust-Region (RTR) method, which utilizes both first- and second-order information of \( f(x) \), ensuring local superlinear convergence. Furthermore, due to its use of second-order curvature information, the RTR algorithm is guaranteed to converge to a global maximizer (set) from any initialization in all trivial maximization problems. This includes initializations lying in the measure-zero set of critical points (such as strict minima or saddle points) where Riemannian gradient ascent fails to escape, assuming a second-order critical step is taken by the subproblem solver.
	
	Before the derivation of the RTR method, key geometric components are introduced. Following \cite{absil_optimization_2008}, the Riemannian metric on \( \mathcal{S} \) is \( \langle a,b \rangle_x = a^T b \) restricted to the tangent space \( \mathcal{T}_x\mathcal{S} := \{z \mid z = P_x y, \forall y \in \mathbb{R}^q\} \), where \( P_x = I_q - xx^T \) is the orthogonal projection onto the tangent space. The projection onto the normal space at \( x \) is \( P_x^{\bot } = xx^T \). Furthermore, let us define the matrix
	\begin{align}\label{eqn:E}
		E(x):=B-\frac{\phi_b(x) }{\phi_w(x) }W+\phi_w(x) D,
	\end{align}
	whose importance, in addition to compacting the notation, will be clarified in Section~\ref{sec:stateoftheart_methods}. Euclidean gradient and Euclidean Hessian of $f(x)$ are
	\begin{align}
		&f'(x) = \frac{2}{\phi_w(x) }E\left(x\right)x,\label{eq:firstDerivative}\\
		f''(x) = \frac{2E\left(x\right)}{\phi_w(x) }+&\frac{8\phi_b(x)Wxx^T W }{\phi_w(x)^3 }-\frac{4\left({\mathrm{Bxx}}^T W-{\mathrm{Wxx}}^T B\right)}{{\phi^2_w(x) } }.\label{eq:secondDerivative}
	\end{align}
	This allows for derivation of Riemannian gradient \cite{absil_optimization_2008}
	\begin{align}
		\mathrm{grad}\;f(x)=P_xf'(x)=f'(x)-2\phi_d(x) x
	\end{align}
	and Riemannian Hessian \cite{absil_extrinsic_2013}
	\begin{align}
		\mathrm{Hess}\;f(x)[h] =P_x f''(x)h+A_x \left(h,P_x^{\bot } f'(x) \right),\label{eq:riemannian_hessian}
	\end{align}
	where $A_x (h,P_x^{\bot } f'(x))=-hx^T P_x^{\bot } f' (x)$ is the Weingarten map for unit sphere $\mathcal{S}^{q-1}$. 
	Compact forms of the Riemannian gradient and Hessian are
	\begin{equation}
		\mathrm{grad}\;f\left(x\right)= 2\left(\frac{E\left(x\right)}{\phi_w(x) }-I\phi_d(x) \right)x, \label{eq:riemannian_gradient_compact}
	\end{equation}
	\begin{equation}
		\mathrm{Hess}\;f(x) [h] = P_x \left(E\left(x\right)-\phi_w \phi_d I+\frac{4\phi_b }{\phi_w^2 }{\mathrm{Wxx}}^T W-\frac{2}{\phi_w } \left({\mathrm{Bxx}}^T W-{\mathrm{Wxx}}^T B\right)\right)h \label{eq:riemannian_hessian_compact}
	\end{equation}
	respectively.

	With the prerequisites established, we introduce the Riemannian Trust-Region (RTR) method for solving \eqref{eq:dualOptimizationProblem} through \eqref{eq:RiemannianOptimization}.
	
	RTR iteratively minimizes a local quadratic model
	\begin{align}
		m_{x}(h):= f\left(x\right)+ \langle h, \mathrm{grad}\;f(x)\rangle_x + \tfrac{1}{2}\langle h, \mathrm{Hess}\;f(x)[h] \rangle_x,
	\end{align}
	approximating objective function $f(x)$ within a region around the current point. Radius of the region is trust-region radius $\Delta_{\textrm{TR}} \in [\Delta_{\textrm{TR, min}}, \Delta_{\textrm{TR, max}}]$, defined in $\tangentspace$.
	
	Trust-region subproblems are usually a minimizing problems \cite{steihaug_conjugate_1983,absil_trust-region_2007}. In our case, given $m_{x}(h)$ and Riemannian metric on $\mathcal{S}$, the trust-region subproblem is expressed as
	\begin{align}\label{eq:TrustRegionSubproblem}
		\begin{split}
			h^*:= \operatorname*{argmin}_{h\in\tangentspace}& \quad  h^T g_x + \tfrac{1}{2} h^T H_x(h) \\
			\textrm{s.t.}&\quad \norm{h}<\Delta_{\textrm{TR}},
		\end{split}
	\end{align}
	where $g_x:=-\mathrm{grad}\;f(x)$ and $H_x(h):=-\mathrm{Hess}\;f(x)[h]$.
	
	The (approximate) solution $h^*$ in $\tangentspace$ is then retracted
	\begin{align}\label{eq:retraction}
		\xi \gets \frac{x + h^*}{\|x + h^*\|},
	\end{align}
	generating a candidate point $\xi\in\mathcal{S}$. 
	
	Using the candidate, the agreement between the model and the actual improvement is evaluated via agreement parameter
	\begin{align}\label{eq:agreement}
		\rho := \frac{f(x) - f(\xi)}{m_x(0)-m_x(h^*)}.
	\end{align}
	The trust-region radius is updated based on this parameter. Finally, if the parameter is smaller than threshold $\underline\rho>0$, the model is deemed inaccurate and the candidate is rejected. The complete algorithm is summarized in Algorithm~\ref{alg:RTR}, where $k_{\textrm{max}}^{\textrm{RTR}}$ denotes the maximum number of iterations and $x_0$ is the initial point.
	
	\begin{algorithm}[h]
		\caption{Riemannian trust-region (RTR) \cite{absil_optimization_2008}}\label{alg:RTR}
		\begin{algorithmic}[1]
			\Require $x_0$, $\Delta_{\textrm{TR}}$, $\Delta_{\textrm{TR, min}}$, $\Delta_{\textrm{TR, max}}$, $\underline\rho$, $k_{\textrm{max}}^{\textrm{RTR}}$
			\State $x \gets x_0$
			\For{$k = 1$ to $k_{\textrm{max}}^{\textrm{RTR}}$}
			\State Obtain $h^*$ by (approximately) solving \eqref{eq:TrustRegionSubproblem}.
			\State Obtain $\xi$ by retracting $h^*$ using \eqref{eq:retraction}.
			\State Obtain $\rho$ by evaluating model agreement using \eqref{eq:agreement}.
			
			\If{$\rho < {1}/{4}$}
			\State $\Delta_{\textrm{TR}} \gets \max \left( {\Delta_{\textrm{TR}}}/{4}, \Delta_{\textrm{TR, min}} \right)$
			\ElsIf{$\rho > {3}/{4}$ \textbf{and} $(\Delta_{\textrm{TR}} - \|h^*\|) \leq \delta_0$}
			\State $\Delta_{\textrm{TR}} \gets \min \left( 2  \Delta_{\textrm{TR}}, \Delta_{\textrm{TR, max}} \right)$
			\EndIf
			
			\If{$\rho > \underline\rho$}
			\State $x \gets \xi$
			\EndIf
			\If {$\norm{\mathrm{grad}\;f(x)}  \leq \delta_0$}
			\State \textbf{return} $x$
			\EndIf
			\EndFor
			\State \textbf{return} $x$
		\end{algorithmic}
	\end{algorithm}
	In all algorithms presented herein, $\delta_0\geq0$ denotes a small numerical threshold used to determine if a value is sufficiently close to zero for purposes of numerical stability.
	\begin{algorithm}[h]
		\caption{Truncated conjugate Gradient (tCG) \cite{steihaug_conjugate_1983}} \label{alg:TcGD}
		\begin{algorithmic}[1]
				\Require $\Delta_{\textrm{TR}}$, $H(h)$, $g$
				
				\State  $h \gets \mathbf{0}\in\real^q$
				\State  $r \gets -g$
				
				\For{$k = 1$ to $q$}
				\If{$\norm{g}^2 \leq \delta_0$}
				\State \textbf{return} $h$
				\EndIf
				\If{$r^TH(r) \leq \delta_0$}
				\State  $\alpha \gets$ { $ \frac{-h^T r + \sqrt{(h^T r)^2 -  \norm{r}^2 (\norm{h}^2 - \Delta_{\textrm{TR}}^2)}}{\norm{r}^2}$}
				\State \textbf{return} $h + \alpha r$
				\EndIf
				\State  $\alpha \gets$ { $\tfrac{-g^Tr}{r^TH(r)}$}
				\If{$\norm{h + \alpha r}^2 \geq \Delta_{\textrm{TR}}^2$}
				\State  $\alpha \gets$ { $ \frac{-h^T r + \sqrt{(h^T r)^2 -  \norm{r}^2 (\norm{h}^2 - \Delta_{\textrm{TR}}^2)}}{\norm{r}^2}$}
				\State \textbf{return} $h + \alpha r$
				\EndIf
				\State $h \gets h + \alpha r$
				\State $\gamma \gets \norm{g}^2$
				\State $g \gets g + \alpha H(r)$
				\State $r \gets -g + \frac{\norm{g}^2}{\gamma} r$
				\EndFor
				\State \textbf{return} $h$
			\end{algorithmic}
	\end{algorithm}
	
	For solving the trust-region subproblem \eqref{eq:TrustRegionSubproblem} the well known truncated Conjugate Gradient (tCG) algorithm (Algorithm \ref{alg:TcGD}) was selected  due to its scalability, efficient handling of large-scale problems through Hessian-vector products, and its ability to naturally detect negative curvature while maintaining low memory requirements. The algorithm can be replaced with any alternative, e.g. \cite{yuan_truncated_2000,yang_new_2010,agarwal_adaptive_2021,pei_sequential_2023}. 
	
	
	\section{Proposed Method}\label{sec:proposed_method}
	While RTR guarantees global convergence \cite{zhang_optimizing_2013}, it does not guarantee convergence to the global maximizer of \eqref{eq:RiemannianOptimization} in nontrivial instances. In this section, we introduce our proposed method, which builds on the Probability-One Homotopy Optimization \cite{watson_probability-one_2002,dunlavy_homotopy_2005} and RTR. This method promises reliable convergence to the optimum by constructing a homotopy $h:\mathcal{S} \times [0,1] \rightarrow\real$ between a trivial function $g(x)$ with a known (or trivial) solution $x^*_g$ and the optimized function $f(x)$:
	\begin{align}\label{eq:homotopyGeneral}
		h(x,\lambda) := \lambda f(x) + (1-\lambda)g(x).
	\end{align}
	As homotopy parameter $\lambda\in[0, 1]$ varies from 0 to 1, the function transitions from the trivial function $h(x,0) = g(x)$ to the target function $h(x,1) = f(x)$. During this transition, the solutions corresponding to each intermediate value of $\lambda$ form a continuous solution trajectory that connects the trivial solution $h(x^*_g,0)$ to the solution $h(\hat x, 1)$ of the target function. 
	
	The homotopy optimization (HOM) algorithm \cite{dunlavy_homotopy_2005} operates by incrementally solving the optimization problem
	\begin{align}\label{eq:homotopyOptimizationProblem}
		\begin{split}
			\hat{x}:= \operatorname*{arg\,max}_{x\in\mathcal{S}}  &\quad h(x,\lambda)
		\end{split}
	\end{align}
	while progressively increasing the homotopy parameter $\lambda$. This procedure traces the solution trajectory along the homotopy, ultimately converging to the maximizer of $f(x)$.
	
	Throughout the homotopy optimization process, \eqref{eq:homotopyOptimizationProblem} is always initialized at the previous solution, making RTR well-suited due to its superlinear local convergence. Furthermore, if certain conditions, discussed in Section~\ref{sec:probability_one_conditions} hold, the HOM method reaches $h(\hat x,1)$ with probability one \cite{van_laarhoven_simulated_1987,locatelli_simulated_2000}.
	
	\begin{proposition}\label{prop:dominance}
		Let $x^*$ denote the global maximizer of \eqref{eq:dualOptimizationProblem}. We propose that $x^*$ is predominantly influenced by one of the two components of the objective function, namely
		$$
		g_1(x) := \frac{\phi_b(x)}{\phi_w(x)} \quad \text{or} \quad g_2(x) := \phi_d(x).
		$$
		Accordingly, a homotopy defined such that the dominant term serves as the trivial component yields a continuous path connecting the trivial solution $h(x^*_g,0)$ to the global maximizer $h(x^*,1)$. Since the dominant term is not known a priori, simultaneous tracking of both homotopies
		\begin{align}
			h_1(x,\lambda) &:= \frac{\phi_b(x)}{\phi_w(x)} + \lambda \phi_d(x) \text{, and} \label{eq:homotopyFunctions1} \\
			h_2(x,\lambda) &:= \lambda \frac{\phi_b(x)}{\phi_w(x)} + \phi_d(x), \label{eq:homotopyFunctions2}
		\end{align}
		ensures convergence to the global maximizer.
	\end{proposition}
	Although we do not have a proof that the proposition holds in all instances of \eqref{eq:dualOptimizationProblem}, we did not observe any counterexamples in our extensive experiments with randomly sampled instances. Details and results of the experiments are presented in Section~\ref{sec:application}.
	
	The homotopies \eqref{eq:homotopyFunctions1} and \eqref{eq:homotopyFunctions2} have trivial solutions $x^*_{g1}$ and $x^*_{g2}$, respectively; maximizing $h_1(x,0)$ corresponds to computing the eigenvector of the largest eigenvalue of the matrix pencil $(B, W)$; maximizing $h_2(x,0)$ corresponds to computing the eigenvector of the largest eigenvalue of $D$.
	
	Tracking both solution trajectories is achieved by running two parallel HOM algorithms, with each homotopy initialized with its respective trivial solution. The final solution is selected as the one yielding the highest function value, which leads to Algorithm~\ref{alg:pHOM} with two tunable parameters $k_{\textrm{max}}^{\textrm{pHOM}}$ and $k_{\textrm{max}}^{\textrm{RTR}}$, the latter being relevant only when RTR is used to solve \eqref{eq:homotopyOptimizationProblem}. In general, increasing the values of the parameters reduces risk of numerical failure at the expense of increased computational cost.
	
	\begin{algorithm}[h]
		\caption{Parallel Homotopy Optimization (pHOM)}\label{alg:pHOM}
		\begin{algorithmic}[1]
			\Require $x^*_{g1}$, $x^*_{g2}$, $k_{\textrm{max}}^{\textrm{pHOM}}$
			\State $\hat x_1 \gets x^*_{g1}$
			\State $\hat x_2 \gets x^*_{g2}$
			\For{ $k=1$ to $k_{\textrm{max}}^{\textrm{pHOM}}$}
			\State Update $\hat x_1$ by (approximately) solving \eqref{eq:homotopyOptimizationProblem} with
			\Statex \hspace{1cm}  $h(x)\gets h_1(x)$, $\lambda \gets k/k_{\textrm{max}}^{\textrm{pHOM}}$ and $\hat x_1$ as initial point.
			\State Update $\hat x_2$ by (approximately) solving \eqref{eq:homotopyOptimizationProblem} with
			\Statex \hspace{1cm}  $h(x)\gets h_2(x)$, $\lambda \gets k/k_{\textrm{max}}^{\textrm{pHOM}}$ and $\hat x_2$ as initial point.
			\EndFor
			\If{$f(\hat x_1) > f(\hat x_2)$}
			\State \textbf{return} $\hat x_1$
			\Else
			\State \textbf{return} $\hat x_2$
			\EndIf
		\end{algorithmic}
	\end{algorithm}
	
	\begin{example}\label{ex:HOM}
		In this example, we focus on an instance of \eqref{eq:dualOptimizationProblem}, specified by the matrices $B$, $D$ and $W$ as follows:
		\begin{equation}
			B \!=\!\!\! \begin{bmatrix}
				-1.08 &  -0.10 &   0.43 &   1.20 &  -1.34\\
				-0.10 &  -1.02 &  -0.02 &   0.80 &  -0.31\\
				0.43 &  -0.02 &  -0.79 &  -0.92 &   1.21\\
				1.20 &   0.80 &  -0.92 &  -3.00 &   2.94\\
				-1.34 &  -0.31 &   1.21 &   2.94 &  -4.11
			\end{bmatrix}\!\!\!,\,
			D \!= \begin{bmatrix}
				-1.16 &  -0.58 &   0.22 &   1.29 &  -1.10\\
				-0.58 &  -0.82 &   0.23 &   0.46 &  -1.04\\
				0.22 &   0.23 &  -0.49 &  -0.19 &   0.20\\
				1.29 &   0.46 &  -0.19 &  -1.57 &   1.10\\
				-1.10 &  -1.04 &   0.20 &   1.10 &  -1.77
			\end{bmatrix}\!\!\!,\nonumber
		\end{equation}
		\begin{equation}
			W = \begin{bmatrix}
				1.17 &   0.11 &  -0.13 &  -0.95 &   0.09\\
				0.11 &   2.54 &  -0.19 &  -0.05 &   0.81\\
				-0.13 &  -0.19 &   1.11 &   0.31 &  -1.46\\
				-0.95 &  -0.05 &   0.31 &   1.34 &  -0.39\\
				0.09 &   0.81 &  -1.46 &  -0.39 &   2.67
			\end{bmatrix}.\nonumber
		\end{equation}
		
		An analysis of this instance shows that there exists at least one local nonglobal maximizer  
		\begin{align}
			\hat x \approx [-0.848,\, 0.140,\, -0.381,\, -0.060,\, 0.336]^T \textrm{ with } f(\hat x) \approx -0.766, \nonumber
		\end{align}
		and isolated global maximizer
		\begin{align}
			x^* \approx [0.477,\, 0.433,\, -0.378,\, 0.656,\, 0.111]^T \textrm{ with } f(x^*) \approx -0.743. \nonumber
		\end{align} 
		
		\begin{figure}[h]
			\centering
			\input{./Example_HOM.tex}
			\caption{Objective values along the solution trajectories for the homotopies $h_1(x, \lambda)$ and $h_2(x, \lambda)$. As predicted by Proposition~\ref{prop:dominance}, the first trajectory (associated with $h_1(x, \lambda)$) converges to the global maximizer $x^*$ (marked by a red star), whereas the second trajectory (associated with $h_2(x, \lambda)$) converges to a local maximizer $\hat x$ (blue triangle).}
			\label{fig:HOM}
		\end{figure}
		
		Solving this problem using the proposed pHOM algorithm (Algorithm~\ref{alg:pHOM}) involves following the solution trajectories associated with the homotopy functions $h_1(x,\lambda)$ and $h_2(x,\lambda)$. Each trajectory is initialized at a trivial solution, denoted by $x^*_{g1}$ and $x^*_{g2}$ respectively. As the homotopy parameter $\lambda\in[0, 1]$ increases from 0 to 1, a local maximizer is computed at each step by solving the subproblem $\max_{x\in\mathcal{S}} \; h_i(x, \lambda)$ using the Riemannian trust-region (RTR) algorithm. Finally, the solution yielding the higher objective value is selected as the global maximizer.
		
		Figure~\ref{fig:HOM} illustrates that, in accordance with Proposition~\ref{prop:dominance}, the global maximizer is predominantly influenced by one of the components in the homotopy sum. In this instance, the dominant contribution comes from the trivial component $g_1(x)$, whereas the local maximizer is driven by $g_2(x)$. 
	\end{example}
	
	\section{Probability-One Conditions}\label{sec:probability_one_conditions}
	Certain conditions, outlined in \cite{watson_probability-one_2002} and later extended to Riemannian manifolds in \cite{seguin_continuation_2022}, ensure that the solution trajectory from \( h(x^*_g,0) \) to \( h(\hat x,1) \) is followed with probability one. Inability to satisfy these conditions does not imply optimization failure; however, it introduces a nonzero probability of tracking failure.
	
	
	Based on \cite[Theorem~3.1]{seguin_continuation_2022}, those conditions are:
	\begin{align}
		&\textrm{1}\big)\quad\;\;\textrm{rank}\left(\mathrm{Hess}\;h(x,\lambda)\right)=q,       \quad \forall(x,\lambda)\in\mathcal{S} \times [0,1], \nonumber\\[4pt]
		&\textrm{2}\big)\quad\norm{\mathrm{Hess}\, h(x,\lambda)^{-1}\!\!\!\left[\tfrac{\partial}{\partial \lambda}\mathrm{grad}\,h(x,\lambda) \right]}   \!<\!    L,   \forall(x,\lambda)  \in  \mathcal{S} \times [0,1];\nonumber
	\end{align}
	
	where $L>0$ is radius of a geodesically convex ball in the Riemannian manifold $\mathcal{S}$.
	
	The first condition guarantees that the Riemannian Hessian is nondegenerate on the tangent bundle, ensuring the path of critical points can be locally parameterized by $\lambda$. The second condition bounds the variation of critical points with respect to $\lambda$, ensuring that the continuation step remains within a region of injectivity and well-posed correction.
	
	Verifying these conditions directly is often difficult in practice. However, structural properties of the objective function can offer indirect evidence. In particular, $h(x,\lambda)$ retains the form of a generalized Rayleigh quotient plus a quadratic term. As shown in \eqref{eq:riemannian_hessian_compact}, the Riemannian Hessian consists of a symmetric operator acting on the tangent space of the sphere. If this operator remains uniformly positive (or negative) definite on the tangent bundle throughout $\mathcal{S} \times [0,1]$, both conditions are satisfied.
	
	In many practical applications, such as B-GTLS (Section~\ref{sec:application}), the matrices $B$, $W$, and $D$ are typically full rank. As the domain $\mathcal{S}$ is compact, the Riemannian Hessian is often uniformly invertible, and the variation of $\mathrm{grad}\;h\left(x,\lambda \right)$ with respect to $\lambda$ is bounded. These structural properties support the expectation that the conditions hold in standard scenarios.
	
	If the Proposition~\ref{prop:dominance} holds, then satisfying these two conditions guarantees that Algorithm~\ref{alg:pHOM} (pHOM) attains the global maximizer of \eqref{eq:dualOptimizationProblem} with probability one.

	\section{State-of-the Art Methods}\label{sec:stateoftheart_methods}
	This section provides a comprehensive review of all three existing approaches for solving \eqref{eq:dualOptimizationProblem}, with each method examined in detail within its respective subsection.

	\subsection{Starting Point Strategy Approach}
	In~\cite{zhang_optimizing_2013}, the extremum of \eqref{eq:RiemannianOptimization} is thoroughly analyzed. The main result, stated in~\cite[Theorem~3.5]{zhang_optimizing_2013}, establishes that for a global maximizer $x^*$, the pair $(\phi_d(x^*)\phi_w(x^*),\, x^*)$ must be an eigenpair corresponding to the largest eigenvalue of the matrix $E(x^*)$, defined in \eqref{eqn:E}.
	
	Building on this theorem, the authors of~\cite{zhang_optimizing_2013} propose the following iterative procedure for locating the global maximizer:
	First, a candidate maximizer $\hat x$ is obtained using the RTR method. If $\hat x$ fails to satisfy the condition in~\cite[Theorem~3.5]{zhang_optimizing_2013}, it is classified as a local maximizer. A new starting point for RTR is then generated using the starting point strategy (SPS) by combining $\hat x$ with the eigenvector $y$ corresponding to the largest eigenvalue of $E(\hat{x})$. 
	
	The process is repeated iteratively, producing a sequence of improved objective values, until the condition in~\cite[Theorem~3.5]{zhang_optimizing_2013} is eventually satisfied. At that point, the resulting point is declared the global maximizer. For a detailed theoretical foundation and justification of this approach, we refer the reader to the original source.
	
	\begin{algorithm}[h]
		\caption{Consolidated Starting Point Strategy (cSPS) \cite{zhang_optimizing_2013}}\label{alg:StartingPointStrategy}
		\begin{algorithmic}[1]
			\Require $\hat x_0$
			\State $\hat{x}\gets \hat{x}_0$
			\State Obtain $y$, the eigenvector corresponding to the largest eigenvalue of $E(\hat x)$
			\If{$\norm{\hat{x}^T y} > \delta_0$}
			\State \textbf{return} $\hat{x}$
			\EndIf
			
			\If{$(\phi_w(y) - \phi_w(\hat{x}))(\phi_d(y) - \phi_d(\hat{x})) \geq 0$}
			\State \textbf{return} $y$
			\EndIf
			
			\If{$(\phi_w(y) > \phi_w(\hat{x}) \text{ and } \phi_d(y) < \phi_d(\hat{x}))$}
			\State $\gamma \gets y^T D  \hat{x}$
			\State $\phi(\hat{x}) \gets \phi_d(\hat{x})$
			\State $\phi(y) \gets \phi_d(y)$
			\Else
			\State $\gamma \gets y^T W  \hat{x}$
			\State $\phi(\hat{x}) \gets \phi_w(\hat{x})$
			\State $\phi(y) \gets \phi_w(y)$
			\EndIf
			
			\State $\beta \gets \frac{2  \operatorname{sign}(\phi(\hat{x}))  \gamma  \phi(\hat{x})}{\sqrt{\phi(\hat{x})^2  (\phi(\hat{x}) - \phi(y))^2 + 4  \phi(\hat{x})^2  \gamma^2}}$
			\State $\alpha \gets \sqrt{1 - \beta^2}$
			
			\State \textbf{return} $\alpha  \hat{x} + \beta  y$
		\end{algorithmic}
	\end{algorithm}
	
	This approach forms the starting point strategy extension of the RTR in both SPS from \cite{zhang_optimizing_2013} and our consolidated version of it (cSPS), presented as Algorithm~\ref{alg:StartingPointStrategy}. Notably, our formulation alters equations for $\alpha$ and $\beta$. The original SPS algorithm is obtained from Algorithm~\ref{alg:StartingPointStrategy} by rewriting $\alpha$ and $\beta$ according the \cite[Section~3.3]{zhang_optimizing_2013} to 
		\begin{equation}
			\alpha = \frac{-\beta\,\gamma + \sqrt{\left(\beta\,\gamma\right)^2 - \phi(\hat{x})\left(\beta^2\,\phi(y) - \phi(\hat{x})\right)}}{\phi(\hat{x})},
		\end{equation}
		\begin{equation}
			\beta = \frac{2\,\operatorname{sign}(\gamma)\,|\gamma|\,\phi(\hat{x})}{\sqrt{\phi(\hat{x})^2 (\phi(\hat{x}) - \phi(y))^2 + 4 \phi(\hat{x})^2 \gamma^2}}.
		\end{equation}
	The original SPS algorithm requires matrix $W$ to be positive definite, otherwise it may generate an infeasible starting point. Our cSPS version does not require positive definite $W$ and attains lower computational complexity owing to its simpler equations.
	
	The biggest limitation of the starting point strategy approach occurs when the $(\phi_d(\hat{x})\phi_w(\hat{x}),\, \hat{x})$ is an eigenpair corresponding to the largest eigenvalue of $E(\hat{x})$. In this case, both algorithms (SPS and cSPS) halts on local, nonglobal maximizer $\hat{x}$.
	
	
	\begin{example}\label{ex:SPS}
		In this example, we consider an instance of the optimization problem \eqref{eq:dualOptimizationProblem}, using the same matrix configuration $B$, $D$ and $W$ as in Example~\ref{ex:HOM}. As previously established in that example, this setup admits a global maximizer $x^*$ and a local nonglobal maximizer $\hat x$.
		
		The algorithms based on \cite[Theorem~3.5]{zhang_optimizing_2013} rely on the assumption that, if the maximization procedure terminates at a point  $\hat x$ such that the pair $(\phi_d(\hat x)\phi_w(\hat x),\, \hat x)$ forms an eigenpair corresponding to the largest eigenvalue of $E(\hat x)$, then $\hat x$ can be declared the global maximizer, i.e., $\hat x = x^*$. However, in the present instance, this condition is satisfied by both the local maximizer $\hat x$ and the global maximizer $x^*$.
		
		We report the following numerical equalities:
		\[
		\begin{aligned}
			\phi_d(\hat x)\phi_w(\hat x) &\approx -0.517, 
			& \lambda_{\max}\!\left(E(\hat x)\right) &\approx -0.517, 
			& \bigl|\phi_d(\hat x)\phi_w(\hat x)-\lambda^{(\hat x)}_{\max}\bigr| &\le \varepsilon,\\
			\phi_d(x^*)\phi_w(x^*) &\approx -0.509, 
			& \lambda_{\max}\!\left(E(x^*)\right) &\approx -0.509, 
			& \bigl|\phi_d(x^*)\phi_w(x^*)-\lambda^{(*)}_{\max}\bigr| &\le \varepsilon,
		\end{aligned}
		\]
		with, numerical tolerance $\varepsilon=10^{-9}$. 
		
		Consequently, both $\bigl(\phi_d(\hat x)\phi_w(\hat x),\,\hat x\bigr)$ and $\bigl(\phi_d(x^*)\phi_w(x^*),\,x^*\bigr)$ are eigenpairs corresponding to the largest eigenvalues of $E(\hat x)$ and $E(x^*)$, respectively. This demonstrates that the sufficient condition in \cite[Theorem~3.5]{zhang_optimizing_2013} can be satisfied by a nonglobal maximizer, causing such methods to systematically misclassify $\hat x$ as the global solution.

	\end{example}
	
	\subsection{Nonlinear Eigenvalue Problem Approach}
	Works \cite{zhang_self-consistent-field-like_2014} and \cite{zhang_nonlinear_2021} expand on \cite[Theorem~3.5]{zhang_optimizing_2013} by replacing RTR iteration with a Self-Consistent-Field-Like (SCF) iteration to solve the following nonlinear eigenproblem:
	\begin{align}\label{eq:nonlinearEigenproblem}
		E(x)x = \phi_d(x)\phi_w(x)x,
	\end{align}
	where $E(x)$ is defined in \eqref{eqn:E}.
	
	Since SCF may oscillate rather than converge, an eigenvalue-based trust-region extension was introduced in \cite{zhang_self-consistent-field-like_2014} to enhance stability. This algorithm, presented in Algorithm~\ref{alg:TRSCF} (TRSCF), utilize a parameter $\gamma > 0$, whose tuning is intended to damp oscillatory behavior.
	
	\begin{algorithm}[h]
		\caption{Trust-Region Self-Consistent-Field-Like iteration (TRSCF) \cite{zhang_self-consistent-field-like_2014}}\label{alg:TRSCF}
		\begin{algorithmic}[1]
			\Require $x_0$, $\gamma$, $k_{\textrm{max}}^{\textrm{TRSCF}}$
			\State $x\gets x_0$
			\State $E_\rho(x) \gets E(x)$
			\For{$k = 1$ to $k_{\textrm{max}}^{\textrm{TRSCF}}$}
			\State Obtain $y$, the eigenvector corresponding to the largest eigenvalue of $E_\rho(x)$ such that $x^T y \geq 0$			
			
			\If{$f(x) > f(y)$}
			\State $\rho \gets \gamma  \left(\lambda_1\left(E_\rho(x)\right) - \lambda_2\left(E_\rho(x)\right)\right)$
			\Else
			\State $\rho \gets 0$
			\EndIf
			\State $x \gets y$
			\State $E_\rho(x) \gets E(x) + \rho x x^T$
			
			\If{$\norm{E(x)  x - \phi_d(x)\phi_w(x)  x} \leq \delta_0$}
			\State \textbf{break}
			\EndIf
			\EndFor
			\State \textbf{return} $x$
		\end{algorithmic}
	\end{algorithm}
	
	A drawback of this method, similar to the previous approach, is that the solution of \eqref{eq:nonlinearEigenproblem} may still correspond to a local maximizer. Furthermore, the parameter $\gamma$ requires careful instance-specific tuning to mitigate oscillations during the optimization process. In some instances, we were unable to identify any value of $\gamma$ that effectively eliminated the instability, despite extensive tuning. While such a value may exist, the difficulty in finding it highlights a fundamental practical limitation of the method.
	
	\begin{example}\label{ex:TRSCF}
		In this example, we consider the optimization problem \eqref{eq:dualOptimizationProblem} with the same matrices as in Example~\ref{ex:HOM}. We systematically explore the performance of Algorithm~\ref{alg:TRSCF} under various settings of $\gamma$. The algorithm is initialized at the fixed starting point $x_0=-I/q$. The maximum number of iterations is set to $k_{\textrm{max}}^{\textrm{TRSCF}}=30$. The parameter $\gamma$ is varied across a logarithmic scale, ranging from $0.001$ to $100$. 
		
		\begin{figure}[h]
			\centering
%
%
\definecolor{mycolor1}{rgb}{0.00000,0.44700,0.74100}%
\definecolor{mycolor2}{rgb}{0.85000,0.32500,0.09800}%
\definecolor{mycolor3}{rgb}{0.92900,0.69400,0.12500}%
\definecolor{mycolor4}{rgb}{0.49400,0.18400,0.55600}%
\definecolor{mycolor5}{rgb}{0.46600,0.67400,0.18800}%
\definecolor{mycolor6}{rgb}{0.30100,0.74500,0.93300}%
\definecolor{mycolor7}{rgb}{0.63500,0.07800,0.18400}%
\definecolor{mycolor8}{rgb}{0.12941,0.12941,0.12941}%
\begin{tikzpicture}

\begin{axis}[%
width=0.856\textwidth,
height=0.542\textwidth,
at={(0\textwidth,0\textwidth)},
scale only axis,
xmin=0,
xmax=30,
xlabel style={font=\color{mycolor8}},
xlabel={$k$},
ymin=-2,
ymax=-0.6,
ylabel style={font=\color{mycolor8}},
ylabel={$f(x)$},
axis background/.style={fill=white},
axis x line*=bottom,
axis y line*=left,
xmajorgrids,
ymajorgrids,
legend style={legend cell align=left, align=left}
]
\addplot [color=black, dashed, line width=1.5pt]
  table[row sep=crcr]{%
0	-0.743377211854518\\
30	-0.743377211854518\\
};
\addlegendentry{$f(x^*)$}

\addplot [color=mycolor1, line width=1.5pt]
  table[row sep=crcr]{%
0	-1.3587485380117\\
1	-1.00603892994178\\
2	-1.99865575428118\\
3	-1.36350860629946\\
4	-1.93101481047231\\
5	-1.33575191922754\\
6	-1.93771308015722\\
7	-1.33839785036741\\
8	-1.93706118576524\\
9	-1.3381399686885\\
10	-1.93712476965629\\
11	-1.33816511771928\\
12	-1.93711856921365\\
13	-1.33816266525025\\
14	-1.93711917386842\\
15	-1.33816290440977\\
16	-1.93711911490381\\
17	-1.33816288108745\\
18	-1.93711912065391\\
19	-1.33816288336179\\
20	-1.93711912009318\\
21	-1.33816288314\\
22	-1.93711912014785\\
23	-1.33816288316163\\
24	-1.93711912014253\\
25	-1.33816288315952\\
26	-1.93711912014305\\
27	-1.33816288315973\\
28	-1.93711912014299\\
29	-1.33816288315971\\
30	-1.937119120143\\
};
\addlegendentry{$\gamma = 0.001$}

\addplot [color=mycolor2, line width=1.5pt]
  table[row sep=crcr]{%
0	-1.3587485380117\\
1	-1.00603892994178\\
2	-1.99865575428118\\
3	-1.36350118863362\\
4	-1.93101334612359\\
5	-1.33574109084168\\
6	-1.93771106065164\\
7	-1.33838682511951\\
8	-1.93705923516757\\
9	-1.33812896817627\\
10	-1.93712281073251\\
11	-1.33815411415411\\
12	-1.93711661125682\\
13	-1.33815166204406\\
14	-1.93711721580216\\
15	-1.33815190116258\\
16	-1.9371171568497\\
17	-1.33815187784485\\
18	-1.93711716259847\\
19	-1.33815188011868\\
20	-1.93711716203788\\
21	-1.33815187989695\\
22	-1.93711716209254\\
23	-1.33815187991857\\
24	-1.93711716208721\\
25	-1.33815187991647\\
26	-1.93711716208773\\
27	-1.33815187991667\\
28	-1.93711716208768\\
29	-1.33815187991665\\
30	-1.93711716208769\\
};
\addlegendentry{$\gamma = 0.003$}

\addplot [color=mycolor3, line width=1.5pt]
  table[row sep=crcr]{%
0	-1.3587485380117\\
1	-1.00603892994178\\
2	-1.99865575428118\\
3	-1.36347519060319\\
4	-1.93100821362649\\
5	-1.33570314303643\\
6	-1.93770398313798\\
7	-1.33834818699791\\
8	-1.93705239894274\\
9	-1.33809041673232\\
10	-1.93711594536425\\
11	-1.33811555201822\\
12	-1.93710974927213\\
13	-1.33811310116487\\
14	-1.93711035343464\\
15	-1.33811334013993\\
16	-1.93711029452467\\
17	-1.33811331683822\\
18	-1.9371103002688\\
19	-1.3381133191103\\
20	-1.93711029970871\\
21	-1.33811331888876\\
22	-1.93711029976332\\
23	-1.33811331891036\\
24	-1.93711029975799\\
25	-1.33811331890825\\
26	-1.93711029975851\\
27	-1.33811331890846\\
28	-1.93711029975846\\
29	-1.33811331890844\\
30	-1.93711029975846\\
};
\addlegendentry{$\gamma = 0.01$}

\addplot [color=mycolor4, line width=1.5pt]
  table[row sep=crcr]{%
0	-1.3587485380117\\
1	-1.00603892994178\\
2	-1.99865575428118\\
3	-1.3633631288414\\
4	-1.93098608821926\\
5	-1.33553965134285\\
6	-1.93767348718044\\
7	-1.33818171379276\\
8	-1.93702293903332\\
9	-1.33792431682143\\
10	-1.93708636052031\\
11	-1.3379494061676\\
12	-1.93708017891621\\
13	-1.33794696070526\\
14	-1.93708078144102\\
15	-1.33794719906564\\
16	-1.93708072271267\\
17	-1.33794717583255\\
18	-1.93708072843695\\
19	-1.33794717809708\\
20	-1.937080727879\\
21	-1.33794717787636\\
22	-1.93708072793339\\
23	-1.33794717789787\\
24	-1.93708072792808\\
25	-1.33794717789578\\
26	-1.9370807279286\\
27	-1.33794717789598\\
28	-1.93708072792855\\
29	-1.33794717789596\\
30	-1.93708072792855\\
};
\addlegendentry{$\gamma = 0.04$}

\addplot [color=mycolor5, line width=1.5pt]
  table[row sep=crcr]{%
0	-1.3587485380117\\
1	-1.00603892994178\\
2	-1.99865575428118\\
3	-1.36313583691817\\
4	-1.93094120024223\\
5	-1.33520843737384\\
6	-1.93761168779514\\
7	-1.33784442335366\\
8	-1.93696322117055\\
9	-1.33758778191575\\
10	-1.9370263926532\\
11	-1.33761277881302\\
12	-1.93702023995671\\
13	-1.33761034415666\\
14	-1.93702083922125\\
15	-1.3376105812886\\
16	-1.93702078085379\\
17	-1.33761055819231\\
18	-1.93702078653869\\
19	-1.33761056044185\\
20	-1.93702078598499\\
21	-1.33761056022275\\
22	-1.93702078603892\\
23	-1.33761056024409\\
24	-1.93702078603367\\
25	-1.33761056024201\\
26	-1.93702078603417\\
27	-1.33761056024221\\
28	-1.93702078603413\\
29	-1.33761056024219\\
30	-1.93702078603413\\
};
\addlegendentry{$\gamma = 0.1$}

\addplot [color=mycolor6, line width=1.5pt]
  table[row sep=crcr]{%
0	-1.3587485380117\\
1	-1.00603892994178\\
2	-1.99865575428118\\
3	-1.36106117671243\\
4	-1.93053074696815\\
5	-1.3322100597137\\
6	-1.93705112190988\\
7	-1.33478903450432\\
8	-1.93642045927163\\
9	-1.33453918580765\\
10	-1.93648155163918\\
11	-1.33456338306265\\
12	-1.93647563444664\\
13	-1.33456103935075\\
14	-1.93647620757304\\
15	-1.33456126635711\\
16	-1.93647615206133\\
17	-1.33456124436979\\
18	-1.93647615743807\\
19	-1.33456124649943\\
20	-1.93647615691729\\
21	-1.33456124629316\\
22	-1.93647615696773\\
23	-1.33456124631314\\
24	-1.93647615696285\\
25	-1.33456124631121\\
26	-1.93647615696333\\
27	-1.33456124631139\\
28	-1.93647615696327\\
29	-1.33456124631138\\
30	-1.93647615696328\\
};
\addlegendentry{$\gamma = 0.6$}

\addplot [color=mycolor7, line width=1.5pt]
  table[row sep=crcr]{%
0	-1.3587485380117\\
1	-1.00603892994178\\
2	-1.99865575428118\\
3	-1.35280377557437\\
4	-1.92888331281094\\
5	-1.32075628209048\\
6	-1.93489147969778\\
7	-1.32309846146431\\
8	-1.93431277761619\\
9	-1.32287366605148\\
10	-1.93436872849895\\
11	-1.32289538469753\\
12	-1.93436331778779\\
13	-1.32289328437143\\
14	-1.93436384103338\\
15	-1.3228934874837\\
16	-1.93436379043261\\
17	-1.3228934678416\\
18	-1.93436379532599\\
19	-1.3228934697411\\
20	-1.93436379485277\\
21	-1.32289346955741\\
22	-1.93436379489853\\
23	-1.32289346957518\\
24	-1.93436379489411\\
25	-1.32289346957346\\
26	-1.93436379489454\\
27	-1.32289346957363\\
28	-1.93436379489449\\
29	-1.32289346957361\\
30	-1.9343637948945\\
};
\addlegendentry{$\gamma = 2$}

\addplot [color=mycolor1, line width=1.5pt]
  table[row sep=crcr]{%
0	-1.3587485380117\\
1	-1.00603892994178\\
2	-1.99865575428118\\
3	-0.855026671030398\\
4	-1.01934356959447\\
5	-0.907614646556054\\
6	-1.90058102020883\\
7	-1.14965825769436\\
8	-1.89427783021645\\
9	-1.150304666437\\
10	-1.89349726295285\\
11	-1.1513234288627\\
12	-1.89409892405878\\
13	-1.15060941751359\\
14	-1.89367774722464\\
15	-1.15110955103768\\
16	-1.89397272518607\\
17	-1.15075983861369\\
18	-1.89376645254237\\
19	-1.15100465712329\\
20	-1.89391084847044\\
21	-1.1508334107682\\
22	-1.89380984285227\\
23	-1.15095326337411\\
24	-1.89388053351154\\
25	-1.1508694141007\\
26	-1.89383107733878\\
27	-1.15092809180002\\
28	-1.89386568638073\\
29	-1.15088703723007\\
30	-1.89384147155916\\
};
\addlegendentry{$\gamma = 7$}

\addplot [color=mycolor2, line width=1.5pt]
  table[row sep=crcr]{%
0	-1.3587485380117\\
1	-1.00603892994178\\
2	-1.99865575428118\\
3	-1.78072691804658\\
4	-1.28185864288585\\
5	-1.95194763325944\\
6	-1.7035077273548\\
7	-1.25452570160122\\
8	-1.95797310235343\\
9	-1.71437077448187\\
10	-1.25829799457951\\
11	-1.95711714180292\\
12	-1.71283641761086\\
13	-1.25776378736398\\
14	-1.95723853196782\\
15	-1.71305420166404\\
16	-1.25783958309706\\
17	-1.95722131137464\\
18	-1.71302331020731\\
19	-1.25782883131605\\
20	-1.95722375420862\\
21	-1.71302769240187\\
22	-1.25783035652859\\
23	-1.95722340767733\\
24	-1.71302707076164\\
25	-1.25783014016793\\
26	-1.95722345683492\\
27	-1.71302715894515\\
28	-1.25783017086002\\
29	-1.95722344986161\\
30	-1.71302714643578\\
};
\addlegendentry{$\gamma = 27$}

\addplot [color=mycolor3, line width=1.5pt]
  table[row sep=crcr]{%
0	-1.3587485380117\\
1	-1.00603892994178\\
2	-1.99865575428118\\
3	-1.93479492744714\\
4	-1.3392491452488\\
5	-1.93742541061156\\
6	-1.86444345275961\\
7	-1.31155850250352\\
8	-1.94395671742763\\
9	-1.87232320328131\\
10	-1.31455482360867\\
11	-1.94323292887665\\
12	-1.87145171390547\\
13	-1.31422284922292\\
14	-1.94331319315402\\
15	-1.87154838056121\\
16	-1.31425966432377\\
17	-1.94330429275598\\
18	-1.87153766160859\\
19	-1.31425558195788\\
20	-1.94330527971527\\
21	-1.87153885022968\\
22	-1.31425603464889\\
23	-1.94330517027208\\
24	-1.87153871842441\\
25	-1.31425598445032\\
26	-1.94330518240815\\
27	-1.8715387330402\\
28	-1.3142559900168\\
29	-1.94330518106239\\
30	-1.87153873141946\\
};
\addlegendentry{$\gamma = 100$}

\end{axis}
\end{tikzpicture}%
			\caption{Evolution of the objective value of Algorithm~\ref{alg:TRSCF} for different values of $\gamma$ parameter, showing consistent oscillatory behavior across all runs.}
			\label{fig:TRSCF}
		\end{figure}

		The Evolution of the objective value accross different values of $\gamma$ presented in Figure~\ref{fig:TRSCF} suggests that, for certain problem instances (including the one considered here) the oscillatory behavior of the optimization method may persist regardless of the choice of $\gamma$. This observation indicates that Algorithm~\ref{alg:TRSCF} may have limited effectiveness in such instances and might not consistently yield reliable solutions.

	\end{example}
	
	
	
	\subsection{Semidefinite Programming Approach}
	The common approach in \cite{nguyen_maximizing_2016,xia_minimizing_2018,wang_linear-time_2019} constructs a univariate semidefinite programming (SDP)-based relaxation of \eqref{eq:dualOptimizationProblem} and subsequently solves the resulting univariate problem.
	
	In \cite{nguyen_maximizing_2016}, following the univariate optimization problem is solved
	\begin{align}\label{eq:SDP_reform}
		\begin{split}
			\lambda^* = \lambda(\mu^*) := \operatorname*{max}_{x\in\real^n}  &\quad \mu + x^TDx,\\
			\textrm{s.t.}&\quad \norm{x}=1\\
			             &\quad x^T(B-\mu W)x\geq 0
		\end{split}
	\end{align}
	where the parameter $\mu \in [\underline{\mu}, \overline{\mu}]$ is bounded by the smallest and largest generalized eigenvalues of the matrix pencil $(B, W)$. After solving this maximization problem, maximizer $x^*$ is recovered.

	\begin{algorithm}[h]
		\caption{Semidefinite programming approach via Grid Search, Pattern Detection, and Quadratic Fit (SDP) \cite{nguyen_maximizing_2016}}
		\label{alg:MaximizeSDP}
		\begin{algorithmic}[1]
			\Require $m$
			\State Compute eigenvalues $\mu$ from the generalized problem $B v = \mu W v$
			\State $\mu_{\min} \gets \min(\mu)$, $\mu_{\max} \gets \max(\mu)$
			\State Generate: $\{\mu_i\}_{i=1}^m \gets$ \texttt{linspace}$(\mu_{\min}, \mu_{\max}, k)$
			\For{$i = 1$ to $m$}
			\State $\lambda_i \gets \lambda(\mu_i)$
			\EndFor
			\State Identify all three-point patterns $\lambda(\mu_{j-1})\leq \lambda(\mu_{j}) \ge \lambda(\mu_{j+1})$
			\If{no patterns found}
			\State $\mu^* \gets \arg\max\{\lambda_1, \lambda_{m}\}$ \Comment{Fallback to best endpoint}
			\Else
			\For{each detected pattern}
			\State Run QFS on $\lambda \gets [\lambda(\mu_{j-1}),\, \lambda(\mu_{j}),\, \lambda(\mu_{j+1})]$
			\EndFor
			\State $\mu^* \gets \mu$ from QFS with the largest $\lambda(\mu)$
			\EndIf
			\State Obtain $X \gets X(\mu^*)$ from SDP
			\State Compute eigendecomposition $X = V S V^T$
			\State Select eigenvector $v^*$ corresponding to largest eigenvalue $s^*$
			\If{$s^* \leq 0$}
			\State Halt gracefully
			\EndIf
			\State \Return $x^*\gets v^* \sqrt{s^*} / \|v^* \sqrt{s^*}\|$
		\end{algorithmic}
	\end{algorithm}
	
		\begin{algorithm}[h]
		\caption{Quadratic Fit Search (QFS) \cite{antoniou_practical_2007}}
		\label{alg:QFS}
		\begin{algorithmic}[1]
			\Require $k_{\max}^{\textrm{QFS}}$, $\varepsilon_\mu$, $\varepsilon_\lambda$
			\For{$k = 1$ to $k_{\max}$}
			\If{$\mu_3 - \mu_1 \leq \varepsilon_\mu$}
			\State \Return $\mu^* \gets \mu_j$ with largest $\lambda_j$
			\EndIf
			\State Compute $\mu_{\text{new}}$ via quadratic interpolation
			\State $\lambda_{\text{new}} \gets \lambda(\mu_{\text{new}})$
			\If{$\mu_{\text{new}} = \mu_2$}
			\State Perturb $\mu_{\text{new}}$ by $\varepsilon_\mu / 2$ towards larger interval
			\EndIf
			\If{$\mu_{\text{new}} > \mu_2$}
			\If{$\lambda_{\text{new}} \leq \lambda_2$}
			\State $\mu_3, \lambda_3 \gets \mu_{\text{new}}, \lambda_{\text{new}}$
			\Else
			\State $\mu_1, \lambda_1 \gets \mu_2, \lambda_2$
			\State $\mu_2, \lambda_2 \gets \mu_{\text{new}}, \lambda_{\text{new}}$
			\EndIf
			\Else
			\If{$\lambda_{\text{new}} \leq \lambda_2$}
			\State $\mu_1, \lambda_1 \gets \mu_{\text{new}}, \lambda_{\text{new}}$
			\Else
			\State $\mu_3, \lambda_3 \gets \mu_2, \lambda_2$
			\State $\mu_2, \lambda_2 \gets \mu_{\text{new}}, \lambda_{\text{new}}$
			\EndIf
			\EndIf
			\If{$\max(\lambda_1, \lambda_3) - \lambda_2 < \varepsilon_\lambda$}
			\State \Return $\mu^* \gets \mu_j$ with largest $\lambda_j$
			\EndIf
			\EndFor
			\State \Return $\mu^* \gets \mu_j$ with largest $\lambda_j$
		\end{algorithmic}
	\end{algorithm}
	
	Evaulation of \eqref{eq:SDP_reform} is done by solving following SDP relaxation 
	\begin{align}\label{eq:SDPProblem}
		\begin{split}
			\lambda(\mu):= \operatorname*{arg\,max}_{X \succeq 0}  \quad & \operatorname{Tr} \left( (\mu I + D) X \right). \\ 
			\text{s.t.} \quad 
			& \operatorname{Tr}(I X) = 1,\,\\
			& \operatorname{Tr} \left( (B - \mu W) X \right) \geq 0\\
		\end{split}
	\end{align}
	At the end of the optimization, the solution $x^*$ is recovered via identity $ xx^T = X $.
	
	Despite being univariate, the optimization function remains multimodal, potentially leading to convergence to a local maximizer. To address this, a two-stage search scheme is used. First, the interval $ [\underline{\mu}, \overline{\mu}] $ is partitioned into a coarse mesh, resulting in set of $m$ equidistant values $\{\mu_i\}_{i=1}^m$. Then, a quadratic fit search refines the maximizer within the subintervals, where $\lambda(\mu_{i-1})\leq \lambda(\mu_{i}) \ge \lambda(\mu_{i+1})$. The quadratic search is terminated either upon reaching the maximum number of iterations $k_{\max}^{\textrm{QFS}}$, when the width of the $\mu$-interval falls below $\varepsilon_\mu$, or when the maximum deviation in $\lambda$ from the current best value falls below $\varepsilon_\lambda$. After refining all subintervals, the candidate with the largest $\lambda(\mu_i)$ is returned as the sought after maximizer. For detailed description of the algorithm, the reader is referred to \cite[Section~4]{nguyen_maximizing_2016}.
	
	In our implementation, we utilize the quadratic fit search Algorithm~\ref{alg:QFS} from \cite[Algorithm~4.3]{antoniou_practical_2007}. 
	
	The primary drawback of this approach is that each evaluation of $ \lambda(\mu) $ requires solving an SDP problem, which is computationally expensive. Additionally, some problems require a very fine mesh to accurately locate the global maximizer, leading to impractically long runtimes. Furthermore, most SDP solvers are suitable for small- to medium-scale problems. As a result, SDP-based formulations are limited in practice: for large-scale instances, current solvers lack the scalability needed for efficient solution, and for smaller problems, alternative methods often outperform SDP solvers in both speed and robustness.

	\begin{example}
		This example studies the same instance of \eqref{eq:dualOptimizationProblem}, as is studied in Example~\ref{ex:HOM}. As a fist step of solving this instance using the Algorithm~\ref{alg:MaximizeSDP}, we obtain smallest and largest generalized eigenvalues of the matrix pencil $(B, W)$ being $\underline{\mu}\approx-4.9717$ and $\overline{\mu}\approx-0.1142$, respectively. We construct a coarse mesh over the interval $ [\underline{\mu}, \overline{\mu}] $ using $m=10$ uniformly distributed values $\{\mu_i\}_{i=1}^{10}$, following the setting used throughout the original paper~\cite{nguyen_maximizing_2016}. Evaluation of the SDP problem \eqref{eq:SDPProblem} over the mesh reveals that the sequence $\lambda(\mu_i)$ increases monotonically for $i=1$ to $i=9$, but then decreases at $i = 10$, i.e., $\lambda(\mu_9)>\lambda(\mu_{10})$. Consequently, the only subinterval in which the local maximizer condition $\lambda(\mu_{i-1})\leq \lambda(\mu_{i}) \ge \lambda(\mu_{i+1})$ is satisfied occurs at index $i=9$. 
		
		This subinterval is maximized using Algorithm~\ref{alg:QFS}, resulting in local maximizer
		\begin{align}
			\hat x \approx [-0.848,\, 0.140,\, -0.381,\, -0.060,\, 0.336]^T \textrm{ with } f(x^*) \approx -0.766. \nonumber
		\end{align}

		\begin{figure}[h]
			\centering
			\input{./Example_SDP.tex}
			\caption{Visualization of the univariate SDP objective \eqref{eq:SDPProblem} (black line), along with the coarse discretization mesh (dotted line). The global optimum $x^*$ is highlighted as a red star, while the local maximizer $\hat{x}$, returned by Algorithm~\ref{alg:MaximizeSDP}, is shown as a blue triangle.}
			\label{fig:SDP}
		\end{figure}
		
		A detailed view of the identified subinterval of the $\lambda(\mu)$ is shown in Figure~\ref{fig:SDP}. As can be seen, the function is bimodal, and the algorithm converges to the local maximum. While employing a finer mesh in place of the coarse discretization would allow the method to locate the global maximizer, this would substantially increase the already significant computational cost. For comparison, the proposed pHOM algorithm, configured as in Experiment~\ref{ex:performanceComparison} (presented in the following section), returns a global maximizer in 1.0743 milliseconds. In contrast, solving the instance using Algorithm~\ref{alg:MaximizeSDP} with current settings requires 11.3282 seconds and yields only a local maximizer. This corresponds to an increase in computational time by more than four orders of magnitude.
		
	\end{example}

	\section{Numerical Analysis}\label{sec:numerical_analysis}
	This section presents three experimental studies. The first experiment investigates whether algorithms designed to improve upon local maximizers can successfully escape local optima. The second compares the proposed pHOM method against state-of-the-art algorithms on a dataset of nontrivial instances. The third investigates the scalability of the top-performing methods as the problem dimension increases.
	
	A dataset of 23,000 nontrivial $q=5$ dimensional instances of Problem~\ref{prob:problem} was constructed to benchmark state-of-the-art algorithms against the proposed pHOM method. Each instance was generated by sampling $B$, $D$ and $W$ matrices with entries drawn from a Student's t-distribution, followed by symmetrization $B\gets B B^T$, $D\gets D D^T$, $W\gets W W^T$. If resulting $W$ was not positive definite, it was resampled. To systematically evaluate behavior under both positive and negative definiteness of $B$ and $D$, each generated matrix triplet $(B,D,W)$ was used to construct two problem instances: one using the original triplet, and the other using $(-B,-D, W)$.
	
	Since triviality cannot be determined analytically from parameters of \eqref{eq:dualOptimizationProblem}, multiple RTR runs were initialized from a fine grid on the unit sphere; instances where all runs converged to the same solution were deemed trivial and discarded. For the remaining instances, local and global maximizers were stored. Due to the high cost of generating such dataset, only low-dimensional dataset was created. The dataset was generated on an NVIDIA DGX A100 system equipped with two AMD EPYC 7742 CPUs, providing a total of 128 physical cores. The generation procedure was fully parallelized across all available cores using MATLAB's parfor construct from the Parallel Computing Toolbox.
	
	All experiments were executed on an HP EliteBook 845 equipped with an AMD Ryzen 7 7840U processor and 32GB of RAM, using MATLAB R2024b with no parallelization.
	
	\begin{experiment}[Testing Recovery from Local Maxima]
	Several algorithms considered in this paper, namely SPS, cSPS, and TRSCF, rely on the condition stated in \cite[Theorem~3.5]{zhang_optimizing_2013}, which is intended to identify whether a candidate solution corresponds to a local, nonglobal maximizer. When this condition is satisfied, the SPS-based algorithms reliably escape the local maximizer using a starting-point strategy. In contrast, the TRSCF algorithm attempts to directly construct an eigenpair that satisfies the same theoretical condition.
	
	In this experiment, we deliberately initialize each of these algorithms at a known local maximizer to evaluate the practical effectiveness of their respective mechanisms. The aim is to assess whether these strategies consistently lead to convergence toward the global maximizer and, more broadly, to examine how reliably the condition from \cite{zhang_optimizing_2013} distinguishes between local and global solutions. For reference, we also include the RTR algorithm, which does not incorporate any such mechanism and is therefore expected to remain at the local maximizer.
	
	Optimization parameters were set as follows: Trust region paramters are $\delta_0=10^{-5}$, $\underline\rho = 0.1$, $\Delta_{\textrm{TR, max}}=1$, $\Delta_{\textrm{TR, min}}=10^{-10}$ and $\Delta_{\textrm{TR}}$ is initialized with 1;  RTR and TRSCF iterations are limited to \( k_{\textrm{max}}^{\textrm{TRSCF}} = k_{\textrm{max}}^{\textrm{RTR}} = 1000 \) and TRSCF scaling factor is carefully selected as \( \gamma = 5 \) to minimize number of instances with undampened oscillatory behavior.
	
	The results of the experiment are presented in Table~\ref{tab:experiment2}, displaying success rates in obtaining global and local maximizers, failure cases, and mean computational time per instance. It is apparent, that none of the methods reliably escape from the local maximizer. As expected, RTR was unable to escape from the local maximizer and halted on the initiated point in 100\% of the instances. Although the integration of SPS and cSPS into RTR enhances global convergence behavior, the approach still fails to converge to the global maximizer in 71\% of the tested instances. While both starting point strategies achieved identical escape rate, the additional computational cost of the SPS leads to greater computation time. Although the TRSCF scaling factor $\gamma$ is tuned to achieve the least failures, the algorithm fails to converge in 23\% of the instances. 
	
	\begin{table}[h]
		\caption{Performance Comparison of Optimization Methods} \label{tab:experiment2}
		\begin{tabularx}{\textwidth}{r *{6}{>{\centering\arraybackslash}X}}
			Alg.             & Global         & Local           & Fail          & Global           & Local           & Fail             \\
			& [\%]           & [\%]            & [\%]          & [ms]             & [ms]            & [ms]             \\
			\hline           &                &                 &               &                  &                 &                  \\[-1.9ex]
			CSPS             &       28.648   &\textbf{71.352}  &             0 &   \textbf{1.0811}&          0.41464&                 -\\   
			SPS              &       28.648   &\textbf{71.352}  &             0 &   \textbf{1.2342}&          0.44039&                 -\\  
			TRSCF            &       12.843   &            64   &\textbf{23.157}&            8.3957&          0.27491&            80.565\\    
			RTR              &            0   &  \textbf{100}   &             0 &                 -&         0.097306&                 -\\
		\end{tabularx}
	\end{table}
	
	This experiment demonstrates, that while \cite[Theorem~3.5]{zhang_optimizing_2013} provides a sufficient condition for local optimality, it is not a reliable indicator of the global maximizer. Consequently, methods that rely on this theorem may fail in nontrivial instances.
	\end{experiment}	
	
	\begin{experiment}[Performance Comparison on Nontrivial Instances]\label{ex:performanceComparison}
	The second experiment provides a comparative evaluation of all algorithms analyzed in this work on a dataset of nontrivial problem instances. The goal is to assess the practical performance of the algorithms in terms of their ability to identify the global solution, and do so efficiently across a variety of problem settings. 
	
	In addition to benchmarking algorithmic performance, this experiment also serves as an empirical test of Proposition~\ref{prop:dominance}. Since all instances are explicitly constructed to be nontrivial, any failure of the pHOM algorithm to converge to the correct solution would indicate that the Proposition~\ref{prop:dominance} may not hold universally. Conversely, the consistent success of pHOM across these instances provides empirical support for the proposition's practical validity.
		
	Algorithms requiring initialization (RTR, SPS, cSPS, TRSCF) were initiated from both promising starting points $x^*_{g1}$ and $x^*_{g2}$, selecting result with the higher functional value, leading to their parallelized variants (pRTR, pSPS, pcSPS, pTRSCF).
	
	Optimization parameters were set identically to the previous experiment. SDP search interval \([ \underline{\mu}, \overline{\mu} ]\) is partitioned into $m=10$ equidistant segments, and quadratic fit search is terminated after $k_{\max}^{\textrm{QFS}}=100$ steps,   or when width or height intervals falls below $\varepsilon_\mu = 0.01$ or $\varepsilon_\lambda = 1e-16$. Parameters for pHOM were slowly increased from conservatively small values, until 100\% convergence was achieved. In this case, the parameters were selected as  \( k_{\textrm{max}}^{\textrm{pHOM}} = 3 \) homotopy steps with \( k_{\textrm{max}}^{\textrm{RTR}} = 10 \) RTR iterations at each step.  
	
	\begin{table}[h]
		\caption{Performance Comparison of Optimization Methods} \label{tab:comparison}
		\begin{tabularx}{\textwidth}{r *{6}{>{\centering\arraybackslash}X}}
			Alg.            & Global         & Local          & Fail          & Global           & Local           & Fail              \\
		                  	& [\%]           & [\%]           & [\%]          & [ms]             & [ms]            & [ms]              \\
			\hline          &                &                &               &                  &                 &                   \\[-1.9ex]
			\textbf{pHOM}   & \textbf{100}   &          0     &          0    & \textbf{0.8546}  &                -&                 - \\     
			pcSPS           &    98.978      &     1.0217     &          0    &         17.474   &          2.3895 &                 - \\ 
			pSPS            &    98.978      &     1.0217     &          0    &         31.675   &           4.1723&                 - \\  
			pRTR            &    98.152      &     1.8478     &          0    &         2.1367   &          0.81094&                 - \\  
			pTRSCF          &    73.509      &    0.90435     &     25.587    &         5.3798   &           10.945&            89.613 \\  
			SDP             &    25.383      &    0.03913     &     74.578    & \textbf{10714}   &   \textbf{17121}&   \textbf{7252.8} \\  
		\end{tabularx}
	\end{table}
	
	Table~\ref{tab:comparison} presents success rates in obtaining global and local maximizers, failure cases (in most cases, the algorithm failed to converge to a stationary point within maximum iteration steps), and mean computational time per instance. 
	Incorporating cSPS or SPS improves the performance of RTR, however, all three algorithms (pRTR, pSPS and pcSPS) fail to reliably converge to the global maximizer due to misidentification of a local maximizer for the global one.
	The TRSCF algorithm exhibits a high failure rate, primarily due to its reliance on problem-specific tuning of the parameter $\gamma$. 
	While finer mesh could improve SDP's convergence and reduce its high failure rate, however it would further increase its already high computational cost. 
	In contrast, pHOM achieves 100\% convergence to global maximizer with the shortest computation time, demonstrating its superiority over state-of-the-art methods.
	
	The fact that pHOM achieved convergence in all instances provides strong empirical evidence that Proposition~\ref{prop:dominance} either holds universally or that potential counterexamples are sufficiently rare to be negligible in practical applications.
	\end{experiment}
	
	\begin{experiment}[Scalability Analysis]
	To evaluate the scalability of algorithms with increasing dimension of the problem being solved, 2,500 experiments were conducted using randomly generated matrices, sampled and symmetrized as in the previous setup. Unlike before, instances were not filtered for triviality due to enormous computational cost of grid RTR search for large instances. 
	
	Given the computational expense of high-dimensional experiments, the comparison with pHOM was limited to a subset of algorithms. The selected algorithms are pRTR, whose consistently low runtime, surpassed only by pHOM, makes it a natural choice for evaluating scalability, and SDP, chosen for its algorithmic scalability. Methods relying on the eigenstructure of $E(x)$ were excluded, as they systematically misidentify local maxima as global ones by design.
	
	Optimization parameters followed those of the previous experiment, with modifications ensuring convergence up to $q = 64$: pRTR used $k_{\textrm{max}}^{\textrm{RTR}} = 1500$; pHOM employed $k_{\textrm{max}}^{\textrm{pHOM}} = 4$ outer iterations, each with $k_{\textrm{max}}^{\textrm{RTR}} = 100$, except for the final iteration where $k_{\textrm{max}}^{\textrm{RTR}} = 1100$. The maximum number of RTR iterations is therefore identical for both pRTR and pHOM.

	\begin{table}[h]
		\caption{Scalability Analysis of Optimization Methods} \label{tab:scalability}
		\begin{tabularx}{\textwidth}{r l *{6}{>{\centering\arraybackslash}X}}
			Alg.& $q$                         & Global       & Local          & Fail             & Global           & Local           & Fail              \\
			&                                 & [\%]         & [\%]           & [\%]             & [ms]             & [ms]            & [ms]              \\
			\hline   &                        &              &                &                  &                  &                 &                   \\[-1.9ex]
			pHOM     &\multirow{3}{*}{4}      & \textbf{100} &           0    &            0     &        10.107    &               - &                 - \\   
			pRTR     &                        &       99.4   &         0.56   &          0.04    &        10.472    &         0.75734 &            34.898 \\   
			SDP      &                        &       35.68  &            0   &          64.32   &         11261    &               - &             14379 \\  \noalign{\vskip 0.3ex}\Xhline{0.1pt}\noalign{\vskip 0.3ex} 
			pHOM     &\multirow{3}{*}{16}     & \textbf{100} &           0    &            0     &        62.536    &              -  &                 - \\    
			pRTR     &                        &       99.04  &          0.64  &            0.32  &          66.058  &          14.698 &              161.5\\     
			SDP      &                        &       26.64  &             0  &           73.36  &          8527.2  &               - &              10716\\  \noalign{\vskip 0.3ex}\Xhline{0.1pt}\noalign{\vskip 0.3ex}  
			pHOM     &\multirow{3}{*}{64}     & \textbf{100} &            0   &               0  &  \textbf{1697.5} &               - &                 - \\   
			pRTR     &                        &       94.56  &           0    &           5.44   &          2312.7  &               - &             3357.4\\    
			SDP      &                        &       28.64  &           0    &          71.36   & \textbf{6697.3}  &               - &             6981.4    
		\end{tabularx}
	\end{table}

	As apparent from Table~\ref{tab:scalability}, pHOM consistently achieved the highest objective function value throughout the experiment, while the convergence rate of both pRTR and SDP degraded in higher dimensions. This could potentially be mitigated by employing more iterations for pRTR and finer mesh and additional quadratic fitting steps for SDP, such adjustments would further favor the scalability of pHOM. As expected, pHOM scales better than pRTR, while retaining superior convergence.
	\end{experiment}

	\section{Application}\label{sec:application}
	To demonstrate the broader applicability of solving Problem~\ref{prob:problem}, beyond the examples introduced earlier, we present its novel connection to multi-sourced Bayesian errors-in-variables identification (B-GTLS), a Bayesian framework for recursive errors-in-variables estimation under noise in all inputs and outputs, which yields an approximate posterior of the identified parameters. The method has been successively developed in \cite{friml_bayesian_2022,friml_recursive_2023,friml_bayesian_2024}, with the latest extension addressing identification from multiple data sources.
	
	From \cite{friml_bayesian_2024}, the log-likelihood function of multi-source B-GTLS is:
	\begin{align}\label{eq:LikelihoodCompact}
		\ln(\likelihood(\tau|\Omega_1,  \Omega_2,\Sigma_1, \Sigma_2)) \propto \frac{\vartheta^T\Omega_1\vartheta}{\vartheta^T\Sigma_1\vartheta}+\frac{\vartheta^T\Omega_2\vartheta}{\vartheta^T\Sigma_2\vartheta},
	\end{align}
	where symmetric matrix $\Omega_i \in \mathbb{R}^{q\times q}$ encapsulates data from $i$-th source and symmetric positive definite matrix $\Sigma_i\in \mathbb{R}^{q\times q}$ define its measurement noise. The parameter $\vartheta = [\tau^T,\, -1]^T$ is expanded vector of the estimated variables $\tau$.
	
	In \cite{friml_bayesian_2024}, Laplace approximation is utilized and each term of the likelihood \eqref{eq:LikelihoodCompact} is approximated individually. This is however precise only in specific case, when the approximate modes are close enough. In general case, Laplace approximation of the full likelihood should be utilized, which requires finding global maximizer of
	\begin{align}\label{eq:optimizationProblem}
		\begin{split}
			\begin{bmatrix}
				\tau^* \\
				-1
			\end{bmatrix}:= \operatorname*{arg\,max}_{\vartheta}  &\quad \frac{\vartheta^T\Omega_1\vartheta}{\vartheta^T\Sigma_1\vartheta}+\frac{\vartheta^T\Omega_2\vartheta}{\vartheta^T\Sigma_2\vartheta},\\
			\textrm{s.t.}&\quad {\vartheta_q}=-1
		\end{split}
	\end{align}
	where $\vartheta_q$ is the last element of $\vartheta$. 
	
	Since the objective function in \eqref{eq:optimizationProblem} is invariant to nonzero scaling of \( \vartheta \), we define \( x = R^{-1} \vartheta \), where \( R^{-T} R^{-1} = \Sigma_2 \), and impose \( \|x\| = 1 \). As a result, \eqref{eq:optimizationProblem} is reformulated as \eqref{eq:dualOptimizationProblem} with $B = R^T\Omega_1R$, $D = R^T\Omega_2R$ and $W = R^T\Sigma_1 R$. 
	
	After solving the reformulized problem \eqref{eq:dualOptimizationProblem} and obtaining $x^*$, $\tau^*$ can be obtained by utilizing \cite[Theorem~4]{friml_bayesian_2024}: $\tau^* = -\chi_{1}/\chi_{2}$, where $ \chi = Rx^* $ is partitioned as $\chi^T=[\chi_1^T,\, \chi_2]$ and $\chi_2$ is the last element of $\chi$. 
	
	\section{Conclusions}\label{sec:conclusions}
	This work advances the state of the art in globally solving Problem~\ref{prob:problem} by proposing a novel homotopy-based optimization approach. Building on existing results in Riemannian optimization, the proposed pHOM algorithm (Algorithm~\ref{alg:pHOM}) significantly improves convergence to the global maximizer  compared to existing techniques. In extensive numerical experiments, the approach achieved 100\% convergence to the global maximizer with the smallest computational cost. This empirical success is supported by Proposition~\ref{prop:dominance}, which, while not proven to hold universally, provides a theoretical basis for the convergence of the homotopy path to the global maximizer. Moreover, under conditions presented in Section~\ref{sec:probability_one_conditions}, convergence to the global maximizer is guaranteed with probability one. The algorithm builds on the Riemannian trust-region method and employs the truncated conjugate gradient algorithm to solve the subproblem. This design allows for straightforward integration of future improvements in either component.
	
	Other methods established in literature, namely the starting point strategy (SPS) extension of the Riemannian trust region algorithm (RTR), the trust-region self-consistent field-like iteration method (TRSCF), and the semidefinite programming (SDP) reformulation approach, are carefully explained, analyzed, and implemented. Their respective limitations in reliably converging to the global maximizer are discussed and illustrated through targeted examples and experiments. In the case of the SPS method, an improvement is introduced in the form of consolidated $\alpha$ and $\beta$ equations, resulting in broader applicability and reduced computational cost.
	
	All methods are compared in a performance study conducted on a dataset of 23,000 low-dimensional, nontrivial instances. The evaluated methods include pHOM, SDP-based approach and parallelized versions of cSPS, SPS, RTR and TRSCF. Each method is assessed based on its ability to converge to the global maximizer, converge only locally, or fail entirely. For each outcome category, the mean computational time is reported. In this setting, the proposed pHOM algorithm consistently outperformed all competing methods in both reliability and efficiency.
	
	To assess scalability, a subset of the best-performing methods (pHOM, pRTR, and the SDP-based method) were tested on higher-dimensional instances with dimensions 4, 16, and 64. These experiments include both trivial and nontrivial instances, generated using the same procedure as in the low-dimensional case. In all tested dimensions, pHOM maintained 100\% convergence to the global maximizer and achieved the shortest mean computational time, confirming its robustness and efficiency even as problem size increases.
	
	The dataset of 23,000 nontrivial instances used in this study is made publicly available, addressing the lack of existing benchmarks of this type. Since generating such instances involves significant computational effort, the dataset represents a valuable resource for future comparative studies. In addition, code for all methods evaluated in this work is provided to support reproducibility.
	
	The applicability of the proposed pHOM algorithm is further demonstrated through its integration within the multi-source Bayesian Generalized Total Least-Squares framework.
	
	Future work will focus on further theoretical analysis of Proposition~\ref{prop:dominance}, including efforts to establish formal guarantees. In addition, improvements in computational efficiency will be explored by leveraging advances in Riemannian trust-region methods and homotopy-based optimization. Finally, we aim to extend Problem~\ref{prob:problem} to encompass sums of more than two generalized Rayleigh quotients, thereby broadening the scope of applicability.


	\section*{Acknowledgments} 
	This work has been performed in the project RICAIP: Research and Innovation Centre on Advanced Industrial Production that has received funding from the European Union's Horizon 2020 research and innovation programme under grant agreement No 857306 and from Ministry of Education, Youth and Sports under OP RDE grant agreement No CZ.02.1.01/0.0/0.0/17\_043/0010085.
	
	This work was co-funded by the European Union under the project Robotics and advanced industrial production (reg. no. CZ.02.01.01/00/22\_008/0004590).


	%
	%

	
	\bibliographystyle{amsplain}
	\bibliography{MyCollection}

\providecommand{\bysame}{\leavevmode\hbox to3em{\hrulefill}\thinspace}
\providecommand{\MR}{\relax\ifhmode\unskip\space\fi MR }
\providecommand{\MRhref}[2]{%
  \href{http://www.ams.org/mathscinet-getitem?mr=#1}{#2}
}
\providecommand{\href}[2]{#2}
\begin{thebibliography}{10}

\bibitem{absil_trust-region_2007}
P.-A. Absil, C.G. Baker, and K.A. Gallivan, \emph{Trust-{Region} {Methods} on
  {Riemannian} {Manifolds}}, Found Comput Math \textbf{7} (2007), no.~3,
  303--330 (en).

\bibitem{absil_optimization_2008}
P-A Absil, Robert Mahony, and Rodolphe Sepulchre, \emph{Optimization algorithms
  on matrix manifolds}, Princeton University Press, 2008.

\bibitem{absil_extrinsic_2013}
P.~A. Absil, Robert Mahony, and Jochen Trumpf, \emph{An {Extrinsic} {Look} at
  the {Riemannian} {Hessian}}, Geometric {Science} of {Information} (Berlin,
  Heidelberg) (Frank Nielsen and Frédéric Barbaresco, eds.), Springer, 2013,
  pp.~361--368 (en).

\bibitem{agarwal_adaptive_2021}
Naman Agarwal, Nicolas Boumal, Brian Bullins, and Coralia Cartis,
  \emph{Adaptive regularization with cubics on manifolds}, Math. Program.
  \textbf{188} (2021), no.~1, 85--134 (en).

\bibitem{antoniou_practical_2007}
Andreas Antoniou and Wu-Sheng Lu, \emph{Practical {Optimization}: {Algorithms}
  and {Engineering} {Applications}}, Springer Science \& Business Media, March
  2007 (en), Google-Books-ID: 6\_2RhaMFPLcC.

\bibitem{dunlavy_homotopy_2005}
Daniel~M. Dunlavy and Dianne~P. O'Leary, \emph{Homotopy optimization methods
  for global optimization.}, Tech. Report SAND2005-7495, Sandia National
  Laboratories (SNL), Albuquerque, NM, and Livermore, CA (United States),
  December 2005.

\bibitem{eberhard_generalized_2005}
Andrew Eberhard, Nicolas Hadjisavvas, and D.~T. Luc, \emph{Generalized
  {Convexity}, {Generalized} {Monotonicity} and {Applications}: {Proceedings}
  of the 7th {International} {Symposium} on {Generalized} {Convexity} and
  {Generalized} {Monotonicity}}, Springer Science \& Business Media, 2005 (en),
  Google-Books-ID: nSMh3zNBG4MC.

\bibitem{friml_bayesian_2024}
Dominik Friml, \emph{Bayesian {Identification} via {Generalized} {Total}
  {Least} {Squares}: {Applied} to {Linear} {Motor} {Drives}}, Doctoral theses,
  {Dissertations}, Brno University of TechnologyBrno, 2024.

\bibitem{friml_bayesian_2022}
Dominik Friml and Pavel Václavek, \emph{Bayesian inference of total
  least-squares with known precision}, 2022 {IEEE} 61st {Conference} on
  {Decision} and {Control} ({CDC}), IEEE, 2022, pp.~203--208.

\bibitem{friml_recursive_2023}
\bysame, \emph{Recursive {Variational} {Inference} for {Total}
  {Least}-{Squares}}, IEEE Control Systems Letters \textbf{7} (2023),
  2839--2844, Publisher: IEEE.

\bibitem{locatelli_simulated_2000}
M.~Locatelli, \emph{Simulated {Annealing} {Algorithms} for {Continuous}
  {Global} {Optimization}: {Convergence} {Conditions}}, Journal of Optimization
  Theory and Applications \textbf{104} (2000), no.~1, 121--133 (en).

\bibitem{nguyen_maximizing_2016}
Van-Bong Nguyen, Ruey-Lin Sheu, and Yong Xia, \emph{Maximizing the sum of a
  generalized {Rayleigh} quotient and another {Rayleigh} quotient on the unit
  sphere via semidefinite programming}, J Glob Optim \textbf{64} (2016), no.~2,
  399--416 (en).

\bibitem{pei_sequential_2023}
Yonggang Pei, Shaofang Song, and Detong Zhu, \emph{A sequential adaptive
  regularisation using cubics algorithm for solving nonlinear equality
  constrained optimization}, Comput Optim Appl \textbf{84} (2023), no.~3,
  1005--1033 (en).

\bibitem{primolevo_towards_2006}
G.~Primolevo, O.~Simeone, and U.~Spagnolini, \emph{Towards a joint optimization
  of scheduling and beamforning for {MIMO} downlink}, 2006 {IEEE} {Ninth}
  {International} {Symposium} on {Spread} {Spectrum} {Techniques} and
  {Applications}, August 2006, ISSN: 1943-7447, pp.~493--497.

\bibitem{steihaug_conjugate_1983}
Trond Steihaug, \emph{The {Conjugate} {Gradient} {Method} and {Trust} {Regions}
  in {Large} {Scale} {Optimization}}, SIAM J. Numer. Anal. \textbf{20} (1983),
  no.~3, 626--637, Publisher: Society for Industrial and Applied Mathematics.

\bibitem{seguin_continuation_2022}
Axel Séguin and Daniel Kressner, \emph{Continuation {Methods} for {Riemannian}
  {Optimization}}, SIAM J. Optim. \textbf{32} (2022), no.~2, 1069--1093,
  Publisher: Society for Industrial and Applied Mathematics.

\bibitem{van_laarhoven_simulated_1987}
Peter J.~M. van Laarhoven and Emile H.~L. Aarts, \emph{Simulated annealing},
  Simulated {Annealing}: {Theory} and {Applications} (Peter J.~M. van Laarhoven
  and Emile H.~L. Aarts, eds.), Springer Netherlands, Dordrecht, 1987,
  pp.~7--15 (en).

\bibitem{wang_linear-time_2019}
Long-Fei Wang and Yong Xia, \emph{A {Linear}-{Time} {Algorithm} for {Globally}
  {Maximizing} the {Sum} of a {Generalized} {Rayleigh} {Quotient} and a
  {Quadratic} {Form} on the {Unit} {Sphere}}, SIAM J. Optim. \textbf{29}
  (2019), no.~3, 1844--1869, Publisher: Society for Industrial and Applied
  Mathematics.

\bibitem{watson_probability-one_2002}
Layne~T. Watson, \emph{Probability-one homotopies in computational science},
  Journal of Computational and Applied Mathematics \textbf{140} (2002), no.~1,
  785--807.

\bibitem{xia_minimizing_2018}
Yong Xia, Longfei Wang, and Shu Wang, \emph{Minimizing the sum of linear
  fractional functions over the cone of positive semidefinite matrices:
  {Approximation} and applications}, Operations Research Letters \textbf{46}
  (2018), no.~1, 76--80.

\bibitem{yang_new_2010}
Yueting Yang, Wenyu Li, and Jing Gao, \emph{A {New} {Conjugate} {Gradient}
  {Trust} {Region} {Method} and its {Convergence}}, 2010 {Third}
  {International} {Joint} {Conference} on {Computational} {Science} and
  {Optimization}, vol.~2, May 2010, pp.~38--41.

\bibitem{yuan_truncated_2000}
Y.~Yuan, \emph{On the truncated conjugate gradient method}, Math. Program.
  \textbf{87} (2000), no.~3, 561--573 (en).

\bibitem{zhang_optimizing_2013}
Lei-Hong Zhang, \emph{On optimizing the sum of the {Rayleigh} quotient and the
  generalized {Rayleigh} quotient on the unit sphere}, Comput Optim Appl
  \textbf{54} (2013), no.~1, 111--139 (en).

\bibitem{zhang_self-consistent-field-like_2014}
\bysame, \emph{On a self-consistent-field-like iteration for maximizing the sum
  of the {Rayleigh} quotients}, Journal of Computational and Applied
  Mathematics \textbf{257} (2014), 14--28.

\bibitem{zhang_nonlinear_2021}
Lei-Hong Zhang and Rui Chang, \emph{A {Nonlinear} {Eigenvalue} {Problem}
  {Associated} with the {Sum}-of-{Rayleigh}-{Quotients} {Maximization}}, CSIAM
  Transactions on Applied Mathematics \textbf{2} (2021), no.~2, 313--335,
  Publisher: Global Science Press.

\end{thebibliography}

\end{document}